\documentclass[11pt, letterpaper]{article}
\usepackage[utf8]{inputenc}
\usepackage{fullpage,times}
\usepackage[T1]{fontenc}

\usepackage{caption}
\usepackage[linesnumbered,algoruled,boxed,lined,algo2e]{algorithm2e}
\usepackage{algorithm}

\RestyleAlgo{ruled}
\SetKwInOut{Parameter}{Parameters}
\SetKwComment{Comment}{/* }{ */}

\usepackage[T1]{fontenc}
\usepackage{amsthm,amsmath,amssymb,amsfonts,authblk,hyperref,setspace}
\usepackage{graphicx}
\usepackage{caption}
\usepackage{subcaption}
\usepackage{xcolor}
\usepackage{authblk}
\usepackage{enumitem}
\usepackage{stfloats}
\usepackage{url}
\usepackage{float}
\usepackage{tikz}

\usepackage{calc}
\usepackage{tikz}
\usepackage{hyperref}
\usepackage{graphicx}
\usepackage{tabularx}

\newtheorem{theorem}{Theorem}[section]

\newtheorem{proposition}[theorem]{Proposition}

\newtheorem{corollary}[theorem]{Corollary}

\DeclareMathOperator*{\ED}{ED}
\DeclareMathOperator*{\Ham}{Ham}

\newcommand\eps{\varepsilon}

\newcommand{\OO}{{\widetilde{O}}}

\newcommand{\N}{{\mathbb{N}}}

\newcommand{\Enc}{{\mathrm{Enc}}}

\newcommand{\MI}{{\mathrm{MIS}}}

\newcommand{\UpdateAG}{{\mathrm{UpdateActiveGrammars}}}

\newcommand{\poly}{{\mathrm{poly}}}

\newcommand{\eval}{{\mathrm{eval}}}

\title{Streaming $k$-edit approximate pattern matching via string decomposition} %TODO Please add

\author[1]{Sudatta Bhattacharya\thanks{Email: sudatta@iuuk.mff.cuni.cz. Partially supported by the Grant Agency of the Czech Republic under the grant agreement no. 19-27871X.}}
\author[1]{Michal Kouck{\'{y}}\thanks{Email: koucky@iuuk.mff.cuni.cz. Partially supported by the Grant Agency of the Czech Republic under the grant agreement no. 19-27871X.}}
\affil[1]{Computer Science Institute of Charles University,
Malostransk{\'e}  n{\'a}m\v{e}st\'{\i} 25,
118 00 Praha 1, Czech Republic}

\begin{document}

\maketitle

\begin{abstract}
In this paper we give an algorithm for streaming $k$-edit approximate pattern matching which uses space $\OO(k^2)$ and time 
    $\OO(k^2)$ per arriving symbol. 
    This improves substantially on the recent algorithm of Kociumaka, Porat and Starikovskaya \cite{kociumaka2022small} which uses space $\OO(k^5)$ and time $\OO(k^8)$ per arriving symbol.
    In the $k$-edit approximate pattern matching problem we get a pattern $P$ and text $T$ and we want to identify all substrings of the text $T$
    that are at edit distance at most $k$ from $P$. 
    In the streaming version of this problem both the pattern and the text arrive in a streaming fashion symbol by symbol and after each symbol of the text we need to report whether there is a current suffix of the text with edit distance at most $k$ from $P$.
    We measure the total space needed by the algorithm and time needed per arriving symbol. 
\end{abstract}

\section{Introduction}

Pattern matching is a classical problem of finding occurrences of a given pattern $P$ in text $T$.
It can be solved in time linear in the size of the pattern and text~\cite{KMP77,boyer1977fast,rabin_karp}.
The classical algorithms use space that is proportional to the pattern size.
In a surprising work~\cite{porat2009exact}, Porat and Porat were the first to design a pattern matching algorithm that uses less space.
They designed an {\em on-line} algorithm that pre-processes the pattern $P$ into a small data structure, and then it 
receives the text symbol by symbol. 
After receiving each symbol of the text, the algorithm is able to report whether the pattern matches the current suffix of the text.
The algorithm uses poly-logarithmic amount of memory for storing the data structure and processing the text.
This represents a considerable achievement in the design of pattern matching algorithms.

Porat and Porat also gave a small-space online algorithm that solves approximate pattern matching up-to Hamming distance $k$, 
{\em $k$-mismatch approximate pattern matching}. 
In this problem we are given the pattern $P$ and a parameter $k$, and we should find all substrings of the text $T$ 
that are at Hamming distance at most $k$ from $P$.
Their algorithm uses $\OO(k^3)$ space, and requires $\OO(k^2)$ time per arriving symbol of the text.
Subsequently this was improved to space $\OO(k)$ and time $\OO(\sqrt{k})$~\cite{clifford2016k}. There has been a series of works \cite{breslauer2014real,clifford2015dictionary,clifford2016k,golan2017real,golan2018towards,gawrychowski2019streaming,starikovskaya2017communication,radoszewski2020streaming,golan_et_al:LIPIcs:2020:12140} on online and streaming pattern matching, 
and the line of work culminated in the work of Clifford, Kociumaka and Porat \cite{rollinghashSODA2019} 
who gave a fully {\em streaming} algorithm with similar parameters as~\cite{clifford2016k}.

In the streaming setting, also the pattern arrives symbol by symbol and we do not have the space to store all of it at once.
An important feature of the algorithm of Clifford, Kociumaka and Porat is that their algorithm not only reports 
the $k$-mismatch occurrences of the pattern but for each $k$-mismatch occurrence of $P$ it can also output the full information about the difference
between $P$ and the current suffix of the text, so called {\em mismatch information}.

Beside approximate pattern matching with respect to Hamming distance, researchers also consider approximate pattern matching with respect to 
other similarity measures such as edit distance. 
Edit distance $\ED(x,y)$ of two strings $x$ and $y$ is the minimum number of insertions, deletions and substitutions needed to transform $x$ into $y$. 
The {\em $k$-edit approximate pattern matching problem} is a variant of the approximate pattern matching where we should find all substrings of $T$
that are at edit distance at most $k$ from $P$.
Since there could be quadratically many such substrings, we usually only require to report for each position in $T$ whether there is a substring
of $T$ ending at that position that has edit distance at most $k$ from $P$.
In the streaming version of the problem we want to output the minimal distance of $P$ to a current suffix of the text after receiving each
symbol of $T$. 
Again we assume that the text as well as the pattern arrive symbol by symbol, and we are interested in how much space the algorithm uses,
and how much time it takes to process each symbol.

Starikovskaya~\cite{starikovskaya2017communication} proposed a streaming algorithm for the $k$-edit pattern matching problem, which uses $\OO(k^{8}\sqrt{m})$ space and takes $\OO(k^{2}\sqrt{m}+k^{13})$ time per arriving symbol. Here, we denote $m=|P|$ and $n=|T|$. 
Recently, using a very different technique Kociumaka, Porat and Starikovskaya \cite{kociumaka2022small} constructed a streaming algorithm, which uses $\OO(k^5)$ space and $\OO(k^8)$ amortized time per arriving symbol of the text. 

In this work we substantially improve on the result of Kociumaka, Porat and Starikovskaya.
We give a streaming algorithm for $k$-edit approximate pattern matching that uses $\OO(k^2)$ space and $\OO(k^2)$ time per arriving symbol.
% Our algorithm is also able to output the edit mismatch information for a minimal occurrence of the pattern at each ending position of the text
    % where $k$-edit occurrence of the pattern appears. (We believe we could tweak the algorithm to output all possible $k$-edit occurrences of the pattern.)
%% MK: Perhaps we could output all the possible occurrences...

\begin{theorem}
Given integer $k\ge 0$, there exists a randomized streaming algorithm for the $k$-edit approximate pattern matching problem that uses $\OO(k^2)$ bits of space and takes $\OO(k^2)$ time per arriving symbol of the text. 
% For each approximate match of the pattern in the text, the edit mismatch information can be retrieved in $O(k^3)$ time as well.
\end{theorem}

% The current bottleneck in our time complexity per arriving symbol is finding prefixes with minimum edit distance of two strings represented by small grammars.
% (See Corollary~\ref{p-edgrammar-suffix-minimum}.)
% We believe that the time complexity of that problem can be reduced to $\OO(k^2)$ which would decrease the time complexity of our algorithm
% also to $\OO(k^2)$.
We speculate that some amortization techniques could bring the time complexity of our $k$-edit approximate pattern matching algorithm further down. 
However, it seems unlikely to achieve complexity below $\OO(k)$ per arriving symbol as one could then solve the plain edit distance problem in 
sub-quadratic time contradicting the Strong Exponential Time Hypothesis (SETH) \cite{backurs2015edit}.
It is an interesting open question to achieve smaller space complexity than $\OO(k^2)$. 
Currently, all known sketching techniques for edit distance that people use for $k$-edit approximate pattern matching give sketches of size $\Omega(k^2)$.

The technique of Kociumaka, Porat and Starikovskaya  \cite{kociumaka2022small} for edit distance pattern matching to large extent emulates the inner working of Hamming approximate pattern matching algorithms.
To that effect Kociumaka, Porat and Starikovskaya had to design a {\em rolling} sketch for edit distance where multiple sketches can be ``homomorphically'' combined into one. 
This requires sophisticated machinery.
Here we use a somewhat different approach.
We use a recent locally consistent decomposition of strings which preserves edit distance of Bhattacharya and Kouck\'y \cite{BK}.
The decomposition in essence translates edit distance to Hamming distance. 
Hence, we apply the $k$-mismatch approximate pattern matching algorithm of Clifford, Kociumaka and Porat \cite{rollinghashSODA2019}
on the stream of symbols coming from the decomposition as a black box.
Bhattacharya and Kouck\'y \cite{BK} also constructed a rolling sketch with limited update abilities, namely adding and deleting a symbol.
We do not use that sketch here.

% \subsection{Our techniques}

\subsection{Related work}
%Hamming distance $\Ham(x,y)$ of two strings $x$ and $y$ of same length is the number of positions where $x$ and $y$ differ.

%Edit distance $\ED(x,y)$ of two strings $x$ and $y$ are the minimum number of insertions, deletions and substitutions needed to transform $x$ into $y$. There is an obvious recursive definition of edit distance, which can be translated into a dynamic programming algorithm. This is the classic $O(n^2)$ algorithm known as the Wagner–Fischer algorithm \cite{WF74}.
%The best known upper bound on the running time is by Landau, Myers and Schmidt \cite{LMS98}, who gave an $O(n+k^2)$ time algorithm, where $k=\ED(x,y)$.
%Backurs and Indyk \cite{backurs2015edit} proved that a truly subquadratic time algorithm for computing edit distance would falsify SETH.

%Approximate pattern matching: Given a pattern $P$ of length $m$ and a text $T$ of length $n$, the goal is the find all the occurrences of $P$ in $T$ which match approximately. This approximate matching can be under hamming distance($k$-hamming approximate pattern matching or $k$-mismatch problem) or edit distance($k$-edit approximate pattern matching).

Landau and Vishkin \cite{landau1986efficient} gave the first algorithm for the {\em $k$-mismatch approximate pattern matching} problem which runs in time $O(k(m\log m + n))$ and takes $O(k(m+n))$ amount of space. This was then improved to $O(m\log m + kn)$ time and $O(m)$ space by Galil and Giancarlo \cite{galil1986improved}. Later, Amir, Lewenstein and Porat \cite{amir2004faster} proposed two algorithms running in time $O(n\sqrt{k\log k})$ and $\widetilde{O}(n + k^{3}(n/m))$. The latter was improved by Clifford, Fontaine, Porat, Sach and Starikovskaya \cite{clifford2016k} who gave an $\widetilde{O}(n + k^{2}(n/m))$ time algorithm. Charalampopoulos, Kociumaka and Wellnitz, in their FOCS'20 paper \cite{charalampopoulos2020faster}, also proposed an $\widetilde{O}(n + k^{2}(n/m))$ time algorithm with slightly better $polylog$ factors. An $\widetilde{O}(n+kn/\sqrt{m})$ time algorithm was given by Gawrychowski and Uznański \cite{gawrychowski2018towards}, which showed a nice tradeoff between the $O(n\sqrt{k\log k})$ and $\widetilde{O}(n + k^{2}(n/m))$ running times. Not only that, they also showed that their algorithm is essentially optimal upto $polylog$ factors, by proving a matching conditional lower bound. The $polylog$ factors in the running time were then improved further by a randomized algorithm by Chan, Golan, Kociumaka, Kopelowitz and Porat \cite{chan2020approximating}, with running time $O(n + kn(\sqrt{\log{m}/m}))$. This problem is thus quite well studied.

For the edit distance counterpart of the problem however, there is still a significant gap between the best upper bound and the known conditional lower bound. Landau and Vishkin \cite{landau1989fast} proposed an $O(nk)$ time algorithm for the problem. This algorithm is still the state of the art for larger values of $k$. Cole and Hariharan \cite{cole2002approximate} gave an algorithm running in time $O(n+m+k^{4}(n/m))$(this runs faster if $m\ge k^{3}$). In their unified approach paper \cite{charalampopoulos2020faster}, Charalampopoulos, Kociumaka and Wellnitz also proposed an algorithm running in time $O(n+m+k^{4}(n/m))$. The same authors in their FOCS'22 paper \cite{charalampopoulos2022faster} gave an algorithm running in time $O(n + k^{3.5}\sqrt{\log{m} \log{k}n/m})$, finally improving the bound after 20 years. For the lower bound, Backurs and Indyk \cite{backurs2015edit} proved that a truly subquadratic time algorithm for computing edit distance would falsify SETH. This would imply that an algorithm for the $k$-edit approximate pattern matching which is significantly faster than $O(n+k^{2}(n/m))$ is highly unlikely.

%In the streaming model, the pattern is given but the text appears one symbol at a time. The task is to output the matches on the go as the symbols arrive one at a time.

Online {\em $k$-mismatch approximate pattern matching} problem was first solved by Benny Porat and Ely Porat in 2009 \cite{porat2009exact}. They gave an online algorithm with running time $\widetilde{O}(k^{2})$ and space $\widetilde{O}(k^{3})$ per arriving symbol of the text. Clifford, Fontaine, Porat, Sach and Starikovskaya in their SODA'16 paper \cite{clifford2016k}, improved it to $\widetilde{O}(k^{2})$ space and $O(\sqrt{k}\log{k} + poly(\log(n)))$ time per arriving symbol of the text. Clifford, Kociumaka and Porat \cite{rollinghashSODA2019} proposed a randomized streaming algorithm which uses $O(k\log{(m/k)})$ space and $O(\log{(m/k)}(\sqrt{k\log{k}}+\log^{3}m))$ time per arriving symbol. The space upper bound is optimal up-to logarithmic factors, matching the communication complexity lower bound. All these algorithms use some form of rolling sketch.

In the streaming model, Starikovskaya proposed a randomized algorithm \cite{starikovskaya2017communication} for the $k$-edit approximate pattern matching problem, which takes $O(k^{8}\sqrt{m}\log^{6}m)$ space and $O((k^{2}\sqrt{m}+k^{13})\log^{4}m)$ time per arriving symbol. Kociumaka, Porat and Starikovskaya \cite{kociumaka2022small} proposed an improved randomized streaming algorithm, which takes $\widetilde{O}(k^5)$ space and $\widetilde{O}(k^8)$ amortized time per arriving symbol of the text. 

%In this paper we propose a randomized streaming algorithm for the $k$-edit approximate pattern matching problem, which takes $\widetilde{O}(k^2)$ space and $\widetilde{O}(k^3)$ time per arriving symbol. 

\section{Notations and preliminaries}
% We use the same notation as~\cite{BK}.
We use a standard notation.
For any string $x = x_1x_2x_2\dots x_n$ and integers $p,q$, $x[p]$ denotes $x_p$, $x[p,q]$ represents substring $x' = x_p\dots x_q$ of $x$, and $x[p,q)=x[p,q-1]$.
If $q<p$, then $x[p,q]$ is the empty string $\eps$. $x[p,\dots]$ represents $x[p,|x|]$, where $|x|$ is the length of $x$.
"$\cdot$"-operator is used to denote concatenation, e.g $x\cdot y$ is the concatenation of two strings $x$ and $y$.
% $\Dict(x) = \{x[i,i+1], i \in [n-1]\}$, is the dictionary of string $x$, which stores all pairs of consecutive symbols that appear in $x$.
For strings $x$ and $y$, $\ED(x,y)$ is the minimum number of modifications ({\em edit operations}) required to change $x$ into $y$, where a single modification can be adding a character, deleting a character or substituting a character in $x$.
All logarithms are based-2 unless stated otherwise.
For integers $p>q$, $\sum_{i=p}^{q} a_i=0$ by definition regardless of $a_i$'s.

\subsection{Grammars}\label{s-grammars}

We will use the following definitions from~\cite{BK}. They are taken essentially verbatim.
Let  $\Sigma \subseteq \Gamma$ be two alphabets and $\# \not\in \Gamma$. A {\em grammar} $G$ is a set of {\em rules} of the type $c \rightarrow ab$
or $c \rightarrow a^r$, where $c \in (\Gamma \cup \{\#\}) \setminus \Sigma$, $a,b \in \Gamma$ and $r\in \N$. 
$c$ is the {\em left hand side} of the rule, and $ab$ or $a^r$ is the {\em right hand side} of the rule. 
$\#$ is the starting symbol. The size $|G|$ of the grammar is the number of rules in $G$. 
We only consider grammars where each $a \in \Gamma \cup \{\#\}$ appears on the left hand side of at most one rule of $G$, we call such grammars {\em deterministic}. 
The $\eval(G)$ is the string from $\Sigma^*$ obtained from $\#$ by iterative rewriting of the intermediate results by the rules from $G$. 
If the rewriting process never stops or stops with a string not from $\Sigma^*$, $\eval(G)$  is undefined.
We use $\eval(G_1,G_2,\dots,G_t)$ to denote the concatenation $\eval(G_1)\cdot \eval(G_2)\cdots \eval(G_t)$.
Using a depth-first traversal of a deterministic grammar $G$ we can calculate its {\em evaluation size $|\eval(G)|$} in time $O(|G|)$. 
Given a deterministic grammar $G$ and an integer $m$ less or equal to its evaluation size, we can construct in time $O(|G|)$
another grammar $G'$ of size $O(|G|)$ such that $\eval(G')=\eval(G)[m,\dots]$.
% Furthermore, given deterministic grammars $G_1,\dots,G_s$ we can construct in time $O(\sum_i|G_i|)$
% another grammar $G'$ of size $s-1+\sum_i |G_i|$ such that $\eval(G')=\eval(G_1)\cdots \eval(G_s)$.
$G'$ will use some new auxiliary symbols.

We will use the following observation of Ganesh, Kociumaka, Lincoln and Saha \cite{ED_compressed_string_Soda22}:

\begin{proposition}[\cite{ED_compressed_string_Soda22}]\label{p-edgrammar}
There is an algorithm that on input of two grammars $G_x$ and $G_y$ of size at most $m$ computes the edit distance $k$ of $\eval(G_x)$ and $\eval(G_y)$ in time $O( (m + k^2) \cdot \poly(\log (m+n)))$, where $n=|\eval(G_x)|+|\eval(G_y)|$.
\end{proposition}

We remark that the above algorithm can be made to output also full information about edit operations that transform $\eval(G_x)$ to $\eval(G_y)$.
We will also use the following proposition which can be obtained from Landau-Vishkin algorithm \cite{landau1986efficient} see e.g. a combination of Lemma 6.2 and Theorem 7.13 in \cite{charalampopoulos2020faster}:

\begin{corollary}\label{p-edgrammar-suffix-minimum}
For every pair of grammars $G_x$ and $G_y$ representing strings $x$ and $y$, respectively, and given a parameter $k$
we can find in time $O( (m + k^2) \cdot \poly(\log (m+n)))$, where $n=|x|+|y|$ and $m=|G_x|+|G_y|$, 
the length of a suffix of $x$ with the minimum edit distance to $y$ among all the suffixes of $x$, 
provided that the edit distance of the suffix and $y$ is at most $k$. 
If the edit distance of all the suffixes of $x$ to $y$ is more than $k$ then the algorithm stops in the given time and reports that no suffix was found.
\end{corollary}

%This follows, since the relevant suffixes of $x$ are of length between $|y|-k$ and $|y|+k$, 
%for each of them we can compute its grammar in time $O(m)$ as noted above, and using Proposition~\ref{p-edgrammar}
%determine its edit distance to $y$.
%If the edit distance computation on two of the strings runs for too long, we can stop it as the edit distance will be
%larger than $k$ so we do not care for it.

\section{Decomposition algorithm}\label{s-decomposition}

Bhattacharya and Kouck\'y \cite{BK} give a string decomposition algorithm ({\em BK-decomposition algorithm}) that splits its input string into blocks, each block represented by a small grammar. 
With high probability over the choice of randomness of the algorithm, 
two strings of length at most $n$ and edit distance at most $k$ are decomposed so that the number of blocks is the same 
and at most $k$ corresponding pairs of blocks differ. 
The edit distance between the two strings corresponds to the sum of edit distances of differing pairs of blocks.

More specifically, the BK-decomposition algorithm gets two parameters $n$ and $k$, $k \le n$, and an input $x$.
It selects at random pair-wise independent functions $C_1,\dots, C_L$  and $S$-wise independent functions $H_0, \dots, H_L$
from certain hash families, 
and using those hash functions it decomposes $x$ into blocks, and outputs a grammar for each of the block.
We call the sequence of the produced grammars the {\em BK-decomposition of $x$}.
Here, parameters $L=\lceil \log_{3/2} n \rceil+3$ and $S=O(k \log^3 n \log^* n)$. 
As shown in~\cite{BK}, the algorithm satisfies the following property.

\begin{proposition}[Theorem 3.1~\cite{BK}]\label{t-decomposition}
    Let $x$ be a string of length at most $n$. The BK-decomposition algorithm outputs a sequence of grammars $G_1,\dots,G_s$ such that for $n$ large enough:
    \begin{enumerate}
        \item With probability at least $1-2/n$, $x=\eval(G_1, \dots, G_s)$.
        \item With probability at least $1-2/\sqrt{n}$, for all $i \in \{1,\dots,s\}$, $|G_i| \le S$.        
    \end{enumerate}
    The randomness of the algorithm is over the random choice of functions $C_1,\dots,C_L$ and $H_0,\dots,H_L$.
\end{proposition}

The functions $C_1,\dots, C_L$ can be described using $O(\log^2 n)$ bits in total and the $S$-wise independent functions $H_0, \dots, H_L$ can be described using $O(S \log^2 n)$ bits in total.
We also need the following special case of Theorem~3.12~\cite{BK}. 

\begin{proposition}[Theorem~3.12~\cite{BK}]\label{t-decomposition-rolling}
    Let $u,x,y\in \Gamma^*$ be strings such that $|ux|,|y|\le n$ and $\ED(x,y) \le k$. Let $G^x_1,\dots,G^x_s$ and  $G^y_1,\dots,G^y_{s'}$ 
    be the sequence of grammars output by the BK-decomposition algorithm on input $ux$ and $y$ respectively, using the same choice of random functions
    $C_1,\dots,C_L$ and $H_0,\dots,H_L$. With probability at least $1-1/5$ the following is true: There exist an integer $r \ge 1$, such that
        $$x = \eval(G^x_{s-s'+1})[r,\dots] \cdot \eval(G^x_{s-s'+2}, \dots, G^x_s) \;\;\;\&\;\;\;y=\eval(G^y_{1}, \dots, G^y_{s'}),$$
     and $$\ED(x,y)=\ED(\eval(G^x_{s-s'+1})[r,\dots], \eval(G^y_1)) + \sum_{i = 2}^{s'} \ED(\eval(G^x_{s-s'+i}),\eval(G^y_{i})).$$
\end{proposition}

The grammars for $x$ can be built incrementally.
For a fixed choice of functions $C_i, H_i$, and a string $x$ we say that grammars $G^x_1,\dots,G^x_t$ are {\em definite} 
in its BK-decomposition $G^x_1,\dots,G^x_s$ 
if for any string $z$ and the BK-decomposition $G^{xz}_1,\dots,G^{xz}_{s'}$ of $xz$ obtained using the same functions $C_i, H_i$, 
$G^x_1 = G^{xz}_1$, \dots, $G^x_t = G^{xz}_t$. 
It turns out that all, but $\OO(1)$ last grammars in the BK-decomposition of $x$ are always definite. 
The following claim appears in~\cite{BK}:

\begin{proposition}[Lemma~4.2~\cite{BK}]\label{l-grammarsuffix}
Let $n$ and $k$ be given and $R=O(\log n \log^* n)$ be a suitably chosen parameter. 
Let $x,z\in \Gamma^*$, $|xz|\le n$. Let $H_0,\dots, H_L, C_1,\dots, C_L$ be given. 
Let $G^x_1,G^x_2,\dots,G^x_s$ be the output of the BK-decomposition algorithm on input $x$, and
 $G^{xz}_1,G^{xz}_2,\dots,G^{xz}_{s'}$ be the output of the decomposition algorithm on input $xz$ using the given hash functions. 
\begin{enumerate}
    \item $G^x_i=G^{xz}_i$ for all $i=1\dots,s-R$.
    \item $|x| \le \sum_{i=1}^{\min(s+R,s')} |\eval(G^{xz}_i)| $.
\end{enumerate}
\end{proposition}

The following claim bounds the resources needed to update BK-decomposition of $x$ when we append a symbol $a$ to it.

\begin{proposition}[Theorem~5.1~\cite{BK}]\label{p-grammarsuffix}
    Let $k\le n$ be given and $R=O(\log n \log^* n)$ be a suitably chosen parameter. 
    Let functions $C_1,\dots,C_L$ and $H_0,\dots,H_L$ be given.
    Let $a\in \Sigma$ and $x\in \Sigma^*$ be of length at most $n$, and let $G^x_1,\dots,G^x_s$ be the grammars output by the BK-decomposition algorithm on input $x$ using functions $C_1,\dots,C_L,H_0,\dots,H_L$. $\UpdateAG(G^x_{s-\min(s,R+1)+1},\dots,G^x_s,a)$ outputs a sequence of grammars $G'_1,\dots,$ $G'_{t'}$
    such that $G^x_1,\dots,G^x_{s-\min(s,R+1)}, G'_1,\dots,G'_{t'}$ is the sequence that would be output by the BK-decomposition algorithm on $x\cdot a$  using the same functions $C_1,\dots,C_L, H_0,\dots,H_L$.
    The update algorithm runs in time $\OO(k)$ and outputs $t' \le 4RL$ grammars.
\end{proposition}

\subsection{Encoding a grammar}\label{s-binencoding}

Let $S$ and $M=O(S \log n) = O(k \log^4 n \log^* n)$ be parameters determined by the BK-decomposition algorithm.
\cite{BK} shows that each grammar of size at most $S$ can be encoded as a string of size $M$ over some polynomial-size
alphabet $\{1, \dots, 2\alpha\}$, where the integer $\alpha$ can be chosen so that $2M/\alpha \le 1/n$. 
The encoding $\Enc$ satisfies that if two grammars differ, their encodings differ in every coordinate. 
The encoding is randomized, and one needs $O(\log n)$ random bits to select the encoding function.
The encoding can be calculated in time linear in $M$, and given $\Enc(G)$ we can decode $G$ in time $O(M)$.
The encoding satisfies:

\begin{proposition}\label{l-binenc}
    Let $G,G'$ be two grammars of size at most $S$ output by BK-decomposition algorithm. 
    Let encoding $\Enc$ be chosen at random.
    \begin{enumerate}
        \item $\Enc(G) \in \{1,\dots, 2\alpha\}^M$.
        \item If $G=G'$ then $\Enc(G)=\Enc(G')$.
        \item If $G\neq G'$ then with probability at least $1-(2M/\alpha)$, $\Ham(\Enc(G), \Enc(G'))=M$, that is they differ in every symbol.
    \end{enumerate}
\end{proposition}

\subsection{$k$-mismatch approximate pattern matching}\label{s-hamming-pattern-matching}

Clifford, Kociumaka and Porat~\cite{rollinghashSODA2019} design a streaming algorithm for {\em $k$-mismatch approximate pattern matching} with the following properties.
The algorithm first reads a pattern $P$ symbol by symbol, and then it reads a text $T$ symbol by symbol. 
Upon reading each symbol of the text it reports whether the word formed by the last received $|P|$ symbols of the text are within Hamming distance at most $k$ from the pattern. 
If they are within Hamming distance at most $k$ we can request the algorithm to report the mismatch information between the current suffix of the text and the pattern.
The parameters $k$ and $n$ are given to the algorithm at the beginning, 
where $n$ is an upper bound on the total length of the pattern and the text.
By {\em mismatch information} between two strings $x$ and $y$ of the same length we understand  
$\MI(x,y)=\{(i,x[i],y[i]);$ $i \in \{1,\dots,|x|\} \textit{ and }x[i]\neq y[i]\}$. 
So the Hamming distance of $x$ and $y$ is $\Ham(x,y)=|\MI(x,y)|$. 
Clifford, Kociumaka and Porat~\cite{rollinghashSODA2019} give the following main theorem.

\begin{proposition}[\cite{rollinghashSODA2019}]\label{p-kmispm}
There exists a streaming {\em $k$-mismatch approximate pattern matching} algorithm which 
uses $O(k \log n \log (n/k) )$ bits of space and 
takes $O((\sqrt{k \log k} + \log^3 n) \log (n/k) )$ time per arriving symbol. 
The algorithm is randomised and its answers are correct with high probability, that is it errs with probability inverse polynomial in $n$. 
For each reported occurrence, the mismatch information can be reported on demand in $O(k)$ time.
\end{proposition}

\section{Algorithm overview}

Now we provide the high-level view of how we proceed. 
We will take the pattern $P$ and apply on it the BK-decomposition algorithm. 
That will give us grammars $G^P_1,G^P_2,\dots, G^P_r$ encoding the pattern.
This has to be done incrementally as the symbols of $P$ arrive.
Then we will incrementally apply the BK-decomposition algorithm on the text $T$.

We will not store all the grammars in memory, instead we will use the {\em $K$-mismatch approximate pattern matching algorithm}
of Clifford, Kociumaka and Porat~\cite{rollinghashSODA2019} ({\em CKP-match algorithm}) on the grammars.
Here $K=k\cdot M$, where $M$ is the encoding size of each grammar.
For a suitable parameter $R=\OO(1)$, we will feed the grammars $G^P_1,\dots,G^P_{r-R}$ to the CKP-match algorithm
as a pattern. In particular, we will encode each grammar by the encoding function $\Enc$ from Section~\ref{s-binencoding}, and  we will feed the encoding into the CKP-match algorithm symbol by symbol.

Then as the symbols of the text $T$ will arrive, we will incrementally build the grammars for $T$ 
while maintaining only a small set of {\em active} grammars. Grammars that become {\em definite} will be fed into
the CKP-match algorithm as its input text. 
(Again each one of the grammars encoded by $\Enc$.)
The CKP-match algorithm will report {\em $K$-mismatch} occurrences of our pattern in the text.
Each {\em $K$-mismatch} occurrence corresponds to a match of the pattern grammars to the text grammars, 
with up-to $k$ differing pairs of grammars.
We will recover the differing pairs of grammars and calculate their overall edit distance.
We will combine this edit distance with the edit distance of the last $R$ grammars of the pattern 
from the last $R$ grammars of the text. 
(The last $R$ grammars of the text contain the active grammars which were not fed into the CKP-match algorithm, yet.) 
If the total edit distance of the match does not exceed the threshold $k$, we report it as an $k$-edit occurrence of $P$ in $T$.
If required we can also output the edit operations that transform the pattern into a suffix of $T$. 
(Among the current suffixes of $T$ we pick the one which gives the smallest edit distance from $P$.)

The success probability of our scheme in reporting a particular occurrence of $P$ in $T$ is some constant $\ge 1/2$. 
Thus, we run the processes in parallel $O(\log n)$ times with independently chosen randomness to achieve 
small error-probability. 

We describe our algorithm in more details next.

\section{Description of the algorithm}

Now we describe one run of our algorithm. 
The algorithm receives parameters $n$ and $k$, based on them it sets parameters $L=O(\log n)$, $R=O(\log n \log^* n)$, $S=O(k \log^3 n \log^*n)$, $M=O(k \log^4 n \log^* n$), $K=k\cdot M = O(k^2 \log^4 n \log^* n)$.
Then it chooses at random pair-wise independent functions $C_1,\dots, C_L$  and $S$-wise independent functions $H_0, \dots, H_L$ needed by the BK-decomposition algorithm. It also selects the required randomness for the encoding function $\Enc$. 
It initializes the CKP-match algorithm for {\em $K$-mismatch} approximate pattern matching on strings of length at most $n\cdot M$.

There are two phases of the algorithm. 
In the first phase the algorithm receives a pattern $P$ symbol by symbol and incrementally builds a sequence of grammars $G^P_1,\dots,G^P_r$ representing the pattern $P$. 
All but the last $R$ grammars are encoded using $\Enc$ and sent to our instance of CKP-match algorithm as its pattern (symbol by symbol of each encoding).
In the second phase our algorithm receives an input text $T$ symbol by symbol. 
It will incrementally build a sequences of grammars $G^T_1,G^T_2,\dots$ representing the received text.
Whenever one of the grammars becomes {\em definite} it is encoded by $\Enc$ and sent to our instance of CKP-match algorithm as the next part of its input text (symbol by symbol).

In the first phase, our algorithm uses the procedure given by Proposition~\ref{p-grammarsuffix} to construct the grammars $G^P_1,\dots,G^P_r$ incrementally by adding symbols of $P$. 
The algorithm maintains a buffer of $2R$ {\em active} grammars which are updated by the addition of each symbol. 
Whenever the number of active grammars exceeds $2R$ we encode the {\em oldest} (left-most) grammars that are definite and pass them to our instance of CKP-match algorithm as the continuation of its pattern. 
The precise details of updating the grammars of the pattern are similar to that of updating them for text which we will elaborate on more.
After the input pattern ends, we keep only $R$ grammars $G^P_{r-R+1},\dots,G^P_{r}$, 
and we send all the other grammars to the CKP-match algorithm. 
Then we announce to the CKP-match algorithm the end of its input pattern.
So the CKP-match algorithm received as its pattern encoding of grammars $G^P_1,\dots,G^P_{r-R}$ in this order.
(In the case we end up with fewer than $R+1$ grammars representing $P$ ($r\le R$), we apply a {\em na\"{\i}ve}
pattern matching algorithm without need for the CKP-match algorithm. 
We leave this simple case as an exercise to the reader.)
For the rest of this description we assume that $r > R$. 

In the second phase, the algorithm will receive the input text $T$ symbol by symbol. 
It will incrementally build a sequence of grammars representing the text using the algorithm from Proposition~\ref{p-grammarsuffix}.
We will keep at most $R$ {\em active} grammars $G^a_1,\dots,G^a_t$ on which the algorithm from Proposition~\ref{p-grammarsuffix} will be applied.
The active grammars represent a current suffix of $T$. 
The prefix of $T$ up-to that suffix is represented by grammars  $G^T_1,\dots,G^T_s$ which are definite. 
Out of those definite grammars we will explicitly store only the last $R$ in a buffer, the other grammars will not be stored explicitly.
(They will be used to calculate the current edit distance and to run the update algorithm from Proposition~\ref{p-grammarsuffix}.)
The encoding of all the definite grammars will be fed into the CKP-match algorithm as its input text whenever we detect that a grammar is definite.

\begin{figure}[htp]
    \centering
    \includegraphics[width=\textwidth]{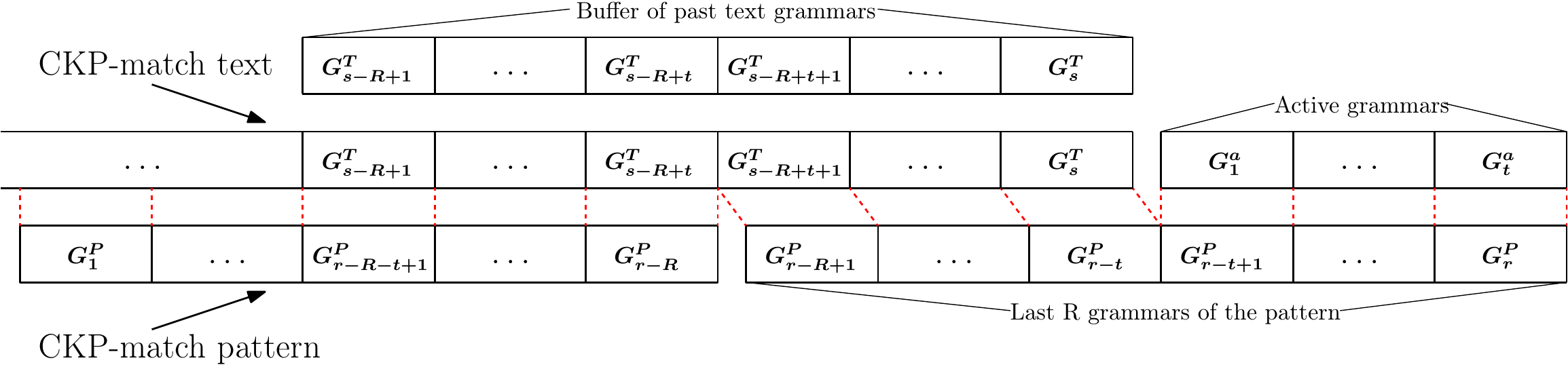}
    \caption{The alignment of text and pattern grammars after arrival of some text symbol. 
    The pattern $P$ is represented by grammars $G^P_1,\dots,G^P_r$. 
    Grammars $G^P_1,\dots,G^P_{r-R}$ are encoded by $\Enc$ and sent to the CKP-match algorithm as its pattern.
    The current text $T$ is represented by the sequence of grammars $G^T_1,\dots,G^T_s,G^a_1,\dots,G^a_t$. 
    Grammars $G^T_1,\dots,G^T_s$ are encoded and committed to the CKP-match algorithm as its text. 
    Grammars $G^a_1,\dots,G^a_t$ are active grammars of the text, and might change as more symbols are added to the text.
    }
    \label{fig-1}
\end{figure}

As the algorithm proceeds over the text it calculates a sequence of integers $m_1,m_2,\dots,m_s$, 
where the algorithm stores only the last $R$ of them in a buffer.
Each value $m_i$ is the minimal edit distance of $\eval(G^P_1, \dots, G^P_{r-R})$ (a prefix of the pattern)
to any suffix of $\eval(G^T_1, \dots, G^T_{i})$ (a suffix of a prefix of the text) if the edit distance is less than $k$. 
$m_i$ is considered infinite otherwise. 
(Values $m_1,\dots,m_{r-R-1}$ are all considered to be infinite.)
The value $m_i$ will be calculated after $G^T_i$ becomes definite and we send the grammar to our CKP-match algorithm.
(The CKP-match algorithm will facilitate its calculation.)
Values $m_i$ will be used to calculate the edit distance of the current suffix of the input text received by the algorithm. See Fig.~\ref{fig-1} for an illustration.

We are ready to describe the basic procedures performed by the algorithm.

\medskip\noindent
{\em Symbol arrival.} Upon receiving the next symbol $a$ of the input text, our algorithm invokes the algorithm from Proposition~\ref{p-grammarsuffix} on the $R+1$ grammars $G^T_{s-R+t},\dots, G^T_s,G^a_1,\dots, G^a_t$ to append the symbol $a$. 
From the algorithm we receive back grammars $G^T_{s-R+t},\dots,G^T_{s},G'^a_1,\dots, G'^a_{t'}$, where $t'<4RL$. 
(Here, $\eval(G'^a_1,\dots, G'^a_{t'})=\eval(G^a_1,\dots, G^a_t) \cdot a$.
The grammars $G^T_{s-R+t},\dots,$ $G^T_{s}$ received from the algorithm are discarded as they are definite and should not change. 
The update algorithm needs them to have the proper context for compression.)
If $t'>R$ then grammars $G'^a_1,\dots, G'^a_{t'-R}$ become definite and we will {\em commit} each of them to the CKP-match algorithm as explained further. 
We will commit them in order $G'^a_1,\dots, G'^a_{t'-R}$.
The remaining grammars $G'^a_{t'-R+1},\dots, G'^a_{t'}$ are relabelled as $G^a_{1},\dots, G^a_{t}$ and become 
the active grammars for the addition of the next symbol.

At this point our algorithm can output the minimal possible edit distance of the pattern to any suffix of the text received up-to this point.
We explain below how such query is calculated.

\medskip\noindent
{\em Committing a grammar.}  When a grammar $G$ becomes definite the algorithm commits the grammar as follows.
Thus far, grammars $G_1,\dots,G_s$ were committed and the sequence of values $m_1,\dots,m_s$ was calculated.
We set $G_{s+1}=G$, calculate encoding $\Enc(G_{s+1})$ and send the encoding symbol by symbol to our CKP-match algorithm.
At this point we can calculate $m_{s+1}$ using the mismatch information provided by our CKP-match algorithm.
If $s+1<r-R$ then we set $m_{s+1}$ to $\infty$ otherwise we continue as follows to calculate $m_{s+1}$.

We query our CKP-match algorithm for the Hamming distance between encoding of $G^P_1,\dots,G^P_{r-R}$ 
(the pattern to the CKP-match algorithm)
and the encoding of $G^T_{s-r+R+2},$ $G^T_{s-r+R+3},\dots, G^T_{s+1}$ (the current suffix of the text of the CKP-match algorithm).
If the Hamming distance is less than $K=k\cdot M$, then we let the CKP-match algorithm to recover the mismatch information.
By the design of the encoding function, if two grammars differ then their encodings differ in all $M$ positions 
(unless the encoding function $\Enc$ fails which happens only with negligible probability.)
Hence, the mismatch information consists of encoding of up-to $k$ pairs of grammars, with their indexes relative to the pattern.
Thus, from the mismatch information we recover pairs of grammars $(G_1,G'_1), \dots, (G_{k'},G'_{k'})$, for some $k'\le k$
where $G_i$ come from the text and $G'_i$ come from the pattern.

If $(G_1,G'_1)$ is not the very first grammar pair $(G^T_{s-r+R+2}, G^P_1)$ (which we recognize by their index in the mismatch information)
then we compute the edit distance for each pair of strings $\eval(G_i)$ and $\eval(G'_i)$, $i=1,\dots, k'$.
We set $m_{s+1}$ to be the sum of those distances.

If $(G_1,G'_1)$ is the pair $(G^T_{s-R+2}, G^P_1)$ then we apply the algorithm from Corollary~\ref{p-edgrammar-suffix-minimum}
to calculate the minimal edit distance between any suffix of $\eval(G_1)$ and the string $\eval(G'_1)$. 
For $i=2,\dots, k'$, we compute the edit distance of $\eval(G_i)$ and $\eval(G'_i)$.
We set $m_{s+1}$ to be the sum of the $k'$ calculated values.

However, if the CKP-match algorithm declares that the Hamming distance of its pattern to its current suffix is more than $K$, we set $m_{s+1}=\infty$. 

Finally, we discard $G_{s-r+R}$ from the buffer of the last $R$ committed grammars, and we discard $m_{s-R+2}$ from the buffer of values $m_i$.
We set $s$ to be $s+1$. This finishes the process of committing a single grammar $G$, and a next grammar might be committed.

\medskip\noindent
{\em Pattern edit distance query.} 
After we process the arrival of a new symbol, update the active grammars as described above and commit grammars as necessary,
the algorithm is ready to answer the edit distance query on the current suffix of the text $T$ and the pattern $P$.
At this point grammars $G^T_1,\dots,G^T_s$ were already committed to the CKP-match algorithm.
There are current active grammars $G^a_1,\dots,G^a_t$ which were not committed to the CKP-match algorithm, and
there are $R$ grammars $G^P_{r-R+1},\dots, G^P_{r}$ of the input pattern that were not committed to the CKP-match algorithm as part of its pattern.
To answer the edit distance query we will compare the edit distance of those last $R$ grammars of pattern $P$ with the last grammars of the text,
and we will combine this with a certain value $m_i$, namely $m_{s-R+t}$.

Let $d=R-t$. If $d>0$, for $i=1,\dots,d$ compute the edit distance of each pair $\eval(G^T_{s-d+i})$ and  $\eval(G^P_{r-R+i})$.
(Each grammar $G^T_{s-d+i}$ is available in the buffer of the last $R$ committed grammars.)
For $i=d+1,\dots,R$, compute the edit distance of each pair $\eval(G^a_{i-d})$ and  $\eval(G^P_{r-R+i})$.
Sum those $R$ values together with $m_{s-d}$. If the sum is less than $k$ output it, otherwise output $\infty$.

Since we are running $O(\log n)$ independent copies of our algorithm, 
each of the copies produces an estimate on the edit distance and we output the smallest estimate. 
That is the correct value with high probability. 

\section{Correctness of the algorithm}

In this section we argue that the algorithm produces a correct output.
First we analyze the probability of certain bad events happening when the algorithm fails and then we argue the correctness of the output assuming none of the bad events happens. 
There are several sources of failure in our algorithm.
\begin{enumerate}
\item The BK-decomposition algorithm might produce a decomposition of either the pattern or some suffix of the text with a grammar that is too big or with grammars that do not represent expected strings. 
(A failure of Proposition~\ref{t-decomposition}.)
\item The BK-decomposition algorithm produces a correct decomposition of the pattern and all suffixes of the text but grammars of some suffix of the text $T$ and the pattern $P$ do not align well. 
(A failure of Proposition~\ref{t-decomposition-rolling}.)
\item The encoding function $\Enc$ fails for some pair of grammars produced by the BK-decomposition algorithm
that the CKP-match algorithm is supposed to compare.
(A failure of Proposition~\ref{l-binenc}.)
\item BK-decomposition algorithm does not fail but the CKP-match algorithm fails to identify a {\em $K$-mismatch} occurrence of its pattern or fails to produce correct mismatch information. 
(A failure of Proposition~\ref{p-kmispm}.)
\end{enumerate}

The failure probability of events 1), 3) and 4) will be each bounded by inverse polynomial in $n$, where $n$ is the parameter sent to those algorithms as an upper bound on the length of the processed strings.
Thus, if we expect our algorithm to process a text and a pattern of size at most $N$, 
we can set the parameter $n$ for the BK-decomposition algorithm to be $N^4$ and for the CKP-algorithm to be $N^4 \cdot M=\OO(N^5)$, 
where $M$ is calculated from $n=N^4$ and $k$ of the BK-decomposition algorithm.
(Parameter $k$  for the BK-decomposition algorithm is set to $k$, and for the CKP-algorithm to $K=k \cdot M=\OO(k^2)$.)
We will run $2\log N$ independent copies of our algorithm on the same text and pattern.
Next we calculate the probability of failure in case 1), 3) and 4) in a particular copy of the algorithm.

\medskip\noindent
{\bf Event 1.} 
There is one pattern $P$ of length at most $N$, the probability of either of the two conditions in Proposition~\ref{t-decomposition} failing on $P$ is at most $4/\sqrt{n} = 4/N^2$. 
The probability of failure of Proposition~\ref{t-decomposition} on any the at most $N$ prefixes of the text $T$
is at most $N\cdot 4/\sqrt{n} = 4/N$. 
Thus the probability of the bad event 1) happening is at most $4/N+4/N^2$.

\medskip\noindent
{\bf Event 3.} 
There are at most $N$ grammars of the pattern encoded by $\Enc$ and there are at most $N$ grammars of the text 
encoded by $\Enc$ and committed. 
Thus there are at most $N^2$ pairs of grammars on which Proposition~\ref{l-binenc} could fail 
by encoding two distinct grammars by strings of Hamming distance less than $M$ (failure in the third part of Proposition~\ref{l-binenc}).
Given our setting of parameters, the probability of the bad event 3) happening is at most $N^2/n = 1/N^2$.

\medskip\noindent
{\bf Event 4.} 
The probability that the CKP-match algorithm fails during its execution is at most $1/n=1/N^4$.

\medskip\noindent
Thus, the probability of a failure of 1), 3) or 4) is at most $5/N$, for $N$ large enough.
We run $2 \log N$ copies of the algorithm so 
the probability that any of the copies fails because of events 1), 3), or 4) is at most $10 \log N/N$.

If none of the events 1), 3) and 4) occurs during the execution of the algorithm then 
the pattern and the text are correctly decomposed into grammars by the BK-decomposition, 
the grammars are properly encoded by $\Enc$, and 
the CKP-match algorithm correctly identifies all the occurrences of the pattern grammars in the committed text grammars, and 
for each of the occurrences we correctly recover the differing pairs of pattern and text grammars.
Assuming this happens, we want to argue that with a high probability our algorithm will correctly identify
$k$-edit occurrences of the pattern $P$ in the text $T$.

After receiving a prefix of the text $T[1,\ell]$, $\ell \le N$, we want to determine whether some suffix  
of $T[1,\ell]$ has edit distance at most $k$ from the pattern $P$.
Let $a$ be such that $T[a,\ell]$ has the minimal distance from $P$. 
Clearly, if the edit distance between $T[a,\ell]$ and $P$ is at most $k$ then $a \in \{\ell - |P|-k+1,\dots, \ell - |P|+k+1\}$.
By Proposition~\ref{t-decomposition-rolling} applied on $u=T[1,a-1]$, $x=T[a,\ell]$ and $y=P$,
each of the $2\log N$ copies of our algorithm has probability at least $4/5$ that 
the grammars of $T$ are well aligned with grammars of $P$.
Being well aligned means that $T[a,\ell]$ is a suffix of $\eval(G^T_{s-r+t+1},\dots, G^T_s, G^a_1,\dots,G^a_t)$
and 
\begin{eqnarray*}
\ED(T[a,\ell], P) &=& \ED(\eval(G^T_{s-r+t+1})[b,\dots], \eval(G^P_1)) \\
&+& \sum_{i=2}^{r-t} \ED(\eval(G^T_{s-r+t+i}),\eval(G^P_{i})) \\
&+& \sum_{i=r-t+1}^r \ED(\eval(G^a_{i-r+t}),\eval(G^P_{i})),
\end{eqnarray*}
for appropriate $b$.
Moreover, the minimality of $a$ implies that 
\begin{eqnarray*}
\ED(T[a,\ell], P) &=& \min_b \ED(\eval(G^T_{s-r+t+1})[b,\dots], \eval(G^P_1)) \\
&+& \sum_{i=2}^{r-t} \ED(\eval(G^T_{s-r+t+i}),\eval(G^P_{i})) \\
&+& \sum_{i=r-t+1}^r \ED(\eval(G^a_{i-r+t}),\eval(G^P_{i})).
\end{eqnarray*}
Notice, regardless of whether Proposition~\ref{t-decomposition-rolling} fails or not, 
the right-hand-side of the last equation is always at least $\ED(T[a,\ell], P)$
since it is an upper-bound on the true edit distance of $P$ to some suffix of $T$.
We will argue that each copy of the algorithm outputs the right-hand-side value of that equation 
if it has value at most $k$, and $\infty$ otherwise.
Moreover, if at least one of the copies of our algorithm has $T[a,\ell]$ and $P$ well aligned, then the minimum
among the values output by the different copies of our algorithm is $\ED(T[a,\ell], P)$.

Since we have $2\log N$ copies of the algorithm, the probability that none of the decompositions aligns $T[a,\ell]$ and $P$ well
is at most $(1/5)^{2\log N} < 1/N^4$. 
This upper-bounds the probability of error of outputting a wrong value of $\min_b \ED(T[b,\ell], P)$ after receiving $\ell$ symbols of the text.
As there will be at most $N$ distinct values of $\ell$, 
the probability of outputting a wrong estimate of the edit distance of $P$ to some suffix of $T$ is at most $N\cdot 1/N^4 = 1/N^3$, 
conditioned on none of the bad events 1), 3) or 4) happening.
Overall, the probability of a failure of our algorithm is at most $O(\log N/N) \le 1/\sqrt{N}$, for $N$ large enough, and it could be made an arbitrary
small polynomial in $N$ by choosing the parameters differently ($n$ vs $N$).

It remains to argue that the copy of our algorithm which aligns $T[a,\ell]$ and $P$ well, outputs their edit distance.
Consider the copy of the algorithm that aligns grammars of $T[a,\ell]$ and $P$ well. 
After arrival of the symbol $T[\ell]$ and updating the grammars, there are active grammars $G^a_1,\dots,G^a_t$, 
committed grammars $G^T_1,\dots,G^T_s$ and the pattern grammars $G^P_1,\dots,G^P_r$. 
If $\ED(T[a,\ell],P)$ is at most $k$ then the number of grammars in which $P$ differs from the last $r$ grammars of $T$ is at most $k$.
Thus the CKP-match algorithm can identify the differing grammars when computing the value $m_{s-R+t}$ 
which is set to 
\begin{eqnarray*}
m_{s-R+t}&=&\min_b \ED(\eval(G^T_{s-r+t+1})[b,\dots], \eval(G^P_1)) \\
&+& \sum_{i=2}^{r-R} \ED(\eval(G^T_{s-r+t+i}),\eval(G^P_{i})).
\end{eqnarray*}
Since, $m_{s-R+t} \le \ED(T[a,\ell],P) \le k$, we have the true value of $m_{s-R+t}$.
Thus, 
\begin{eqnarray*}
\ED(T[a,\ell], P) &=& m_{s-R+t} \\
&+& \sum_{i=r-R+1}^{r-t} \ED(\eval(G^T_{s-r+t+i}),\eval(G^P_{i})) \\
&+& \sum_{i=r-t+1}^r \ED(\eval(G^a_{i-r+t}),\eval(G^P_{i})).
\end{eqnarray*}
That is precisely how we evaluate the edit distance query of our algorithm.

If $\ED(T[a,\ell],P)>k$ then we will output a value $>k$ as we output some upper bound on the edit distance. Any value $>k$ is treated as the infinity.

\section{Time complexity of the algorithm}
In the first phase, we incrementally construct the grammars for the pattern $P$, using the BK-decomposition algorithm from Proposition~\ref{p-grammarsuffix} on each symbol of $P$ at a time. 
Updating the active grammars for each new symbol takes $\OO(k)$ time, committing each of the possible $\OO(1)$ definite grammars to the CKP-match algorithm takes $\OO(M \cdot \sqrt{K})=\OO(k^2)$. Thus the time needed per arriving symbol of the pattern is $\OO(k^2)$. 

For each symbol of the text that arrives during the second phase of the algorithm we need to update the active grammars of the text, update $m_s$, and evaluate the edit distance of the pattern from the current suffix of text.
This includes parts {\em Symbol arrival, Committing a grammar} and {\em Pattern edit distance query} of the algorithm.

{\em Symbol arrival.} Appending a symbol using the BK-decomposition algorithm from Proposition~\ref{p-grammarsuffix} takes $\OO(k)$ time.

{\em Committing a grammar.} Encoding the grammar takes $O(M)$ time using the algorithm from Proposition~\ref{l-binenc}, and committing it
to the CKP-match algorithm takes time $\OO(k^2)$, as in the pattern case. 

Querying the CKP-match algorithm for Hamming distance $K$ takes $O(K)=\OO(k^2)$ time. This recovers at most $k$ pairs of distinct grammars $(G_i,G^{'}_i)$, $1\le i\le k$. Computing edit distance $k_i$ of each pair of strings $\eval(G_i)$ and $\eval(G^{'}_i)$, takes $\OO(S + k_i^2) = \OO(k+k_i^2)$ time using Proposition~\ref{p-edgrammar}. If $\sum_i k_i \le k$, the total time for the edit distance computation is bounded
by $\OO(k^2)$. If the computation runs for longer we can stop it as we know $m_{s}$ is larger than $k$.
Running the algorithm from Corollary~\ref{p-edgrammar-suffix-minimum} on the first pair of distinct grammars to compute the minimum edit distance between any suffix of $\eval(G_1)$ and the string $\eval(G^{'}_1)$ takes $\OO(S + k^2)$ time. 
Thus committing a grammar takes time at most $\OO(k^2)$ where the longest time takes the minimization algorithm on the first pair of grammars.

{\em Pattern edit distance query.} This step requires the alignment of the last $R$ grammars of the pattern with the appropriate grammars of the text and computing their edit distances. Using Proposition~\ref{p-edgrammar}, computing edit distances of $R$ pairs of grammars takes $R\times \OO(k^2) = \OO(k^2)$ time.

As there are at most $\OO(1)$ committed grammars after processing each new symbol, the total time of this step is $\OO(k^2)$ per arriving symbol.

\section{Space complexity of the algorithm}

During either phase of the algorithm, we store $O(RL)=\OO(1)$ active and updated grammars and buffer last $O(R)$ committed grammars. 
This requires space $\OO(k)$. 
Furthermore, the CKP-match algorithm requires $\OO(K)=\OO(k^2)$ space.
The edit distance algorithm of Proposition~\ref{p-edgrammar} cannot use more space than its running time so each invocation uses at most $\OO(k^2)$ space.
Similarly, Corollary~\ref{p-edgrammar-suffix-minimum} uses space $\OO(k^2)$.
Thus our algorithm uses space at most $\OO(k^2)$ at any point during its computation.

\section*{Acknowledgements}

We thank Tomasz Kociumaka for pointing to us references for Corollary~\ref{p-edgrammar-suffix-minimum}. We thank anonymous reviewers for helpful comments.

\bibliographystyle{alpha}
\bibliography{edit-pat-icalp}

\end{document}